\documentclass[prl,twocolumn,superscriptaddress,preprintnumbers,amssymb,epsfigs]{revtex4}

\usepackage{graphicx}
\usepackage{dcolumn}
\usepackage{bm}
\usepackage{wasysym}

\def\beq{\begin{equation}}
\def\eeq{\end{equation}}
\def\beqn{\begin{eqnarray}}
\def\eeqn{\end{eqnarray}}

\def\be{\begin{equation}}
\def\ee{\end{equation}}
\def\ba#1{\begin{array}{#1}}
\def\ea{\end{array}}
\def\bn{\begin{enumerate}}
\def\en{\end{enumerate}}

\def\r{\right}
\def\l{\left}

\def\G{\Gamma}
\def\sz{\hat{\sigma}^{z}}
\def\sx{\hat{\sigma}^{x}}

\def\ra{\lambda}

\begin{document}
\title{Entanglement entropy of random quantum critical points in one dimension}
\author{G. Refael}
\affiliation{Kavli Institute of Theoretical Physics, Santa Barbara, CA 93106}
\author{J.~E.~Moore}
\affiliation{Department of Physics, University of California,
Berkeley, CA 94720}
\affiliation{Materials Sciences Division,
Lawrence Berkeley National Laboratory, Berkeley, CA 94720}
\begin{abstract}
For quantum critical spin chains without disorder, it is known that
the entanglement of a segment of $N \gg 1$ spins with the remainder is
logarithmic in $N$ with a prefactor fixed by the central charge of the
associated conformal field theory.  We show that for a class of
strongly random quantum spin chains, the same logarithmic scaling
holds for mean entanglement at criticality and defines a critical
entropy equivalent to central charge in the pure case.  This effective
central charge is obtained for Heisenberg, XX, and quantum Ising
chains using an analytic real-space renormalization group approach
believed to be asymptotically exact. For these random chains, the
effective universal central charge is characteristic
of a universality class and is consistent with a $c$-theorem.
\end{abstract}
\maketitle

Second-order phase transitions at zero temperature show universal
scaling behavior determined by the collective physics of
quantum fluctuations.  Recently the scaling of the entanglement near
such quantum critical points has been of special interest: at a
quantum critical point, the length scale over which different regions
of the system are entangled becomes divergent.  The entanglement near
criticality was shown to obey a universal scaling law in some
one-dimensional (1D) systems. Most quantum phase transitions in pure 1D
systems are invariant under local conformal transformations,
and the entanglement at a critical point is related to the central
charge of the associated conformal field
theory~\cite{holzhey,vidalent}.

Our primary result is that there exists universal
entanglement scaling even for a class of disordered quantum critical
points in one dimension that are not conformally invariant.
The specific theories considered here describe random quantum spin chains: the
Heisenberg, XX, and quantum Ising chains with random nearest-neighbor
coupling have been previously found,~\cite{DSF94, DSF95} using a real-space
renormalization-group (RG) approach, to be described by strongly disordered
critical points, as reviewed below.

The entanglement of a pure quantum-mechanical state $|\psi\rangle$
with respect to a partition into two subsystems $A$ and $B$ is the von
Neumann entropy of the reduced density matrix for either subsystem:
\beq
S = -{\rm Tr} \rho_A \log_2 \rho_A = - {\rm Tr} \rho_B \log_2 \rho_B
\eeq
where the reduced density matrix $\rho_A$ for subsystem $A$ is
obtained by tracing over a basis $\phi_B^i$ of subsystem $B$
\beq
\langle \phi^1_A | \rho_A | \phi^2_A \rangle = \sum_i (\langle \phi^1_A| \times \langle \phi^i_B|) \psi \rangle \langle \psi (|\phi^2_A \rangle \times |\phi^i_B\rangle).
\eeq
Note that this pure-state entanglement of a spin chain~\cite{vidal}
differs from the two-spin mixed-state entanglement~\cite{osterloh,osborne},
which also has special behavior near a phase transition, but is only
nonzero at short distances and is tied to the spin-spin correlator
rather than the central charge~\cite{kanpur}.

For conformally invariant critical theories in one dimension~\cite{holzhey}, the entanglement of a finite region of size $L$ with the remainder of the system grows logarithmically in $L$ at a critical point, while away from criticality the entanglement is localized near the boundaries of the subsystem and goes to a constant for large $L$.  For critical lattice models like quantum spin chains, the entanglement of a segment of $L$ sites with the remaining sites grows as $\log_2 L$, with a coefficient determined by the central charge of the conformal field theory (CFT)~\cite{vidal}:
\beq
S_N \sim {c+{\bar c} \over 6} \log_2 L.
\label{enteq}
\eeq
Here $c$ and ${\bar c}$ are the holomorphic and antiholomorphic central charges of the 
CFT (for the cases we discuss $c={\bar c}$), which control several physical properties such as low-temperature specific heat.  Slightly off criticality, the spin chain has a finite entanglement length $\xi$, and the entanglement saturates as $L \rightarrow \infty$ to $S \sim {c+{\bar c} \over 6} \log_2 \xi.$


An example of a quantum spin chain that is critical without disorder
is the antiferromagnetic Heisenberg model; the ground state of the spin-half chain
\beq
H = J \sum_i {\bf S}_i \cdot {\bf S}_{i+1} = J \sum_i (S^z_i S^z_j + \frac{S^+_i S^-_j + S^-_i S^+_j}{2}) 
\eeq
is quantum critical for the antiferromagnetic case $J>0$.  Staggered spin-spin correlations, $(-1)^{|i-j|} \langle {\bf S}_i \cdot {\bf S}_j \rangle$, fall off as $1/|i-j|$ up to logarithmic corrections.

The nature of quantum spin chains
with quenched randomness at zero temperature is quite different from
the above pure case.  It is believed that any initial randomness
in the distribution of couplings drives the system at long
distances to a random quantum critical point: in
RG language, disorder is a relevant
perturbation to the pure critical points.  This flow to strong
disorder occurs for the Heisenberg chain, 
the XX chain, which has coupling only in two spin directions,
and the quantum Ising chain, which has couplings in one spin
direction plus a normal magnetic field (both made random).


The low energy properties of the random Heisenberg and XX models are described
by the random singlet phase 
\footnote{In fact, this is true for all anisotropic XXZ spin models $H =
  J \sum_i (\Delta S^z_i S^z_j + \frac{S^+_i S^-_j + S^- i S^+_j}{2})$ for
spin-1/2 sites and with $-0.5<\Delta<1$.}. This is shown using the real-space RG approach
  \cite{MaDas1979,MaDas1980, DSF94}. We review
  the real-space RG approach and the random singlet
  ground state, starting with the random Heisenberg
  Hamiltonian
\beq
H = \sum_i J_i {\bf S}_i \cdot {\bf S}_{i+1}.
\label{Hmodel}
\eeq
The same results apply to the XX chain. In Eq. (\ref{Hmodel}), $J_i$'s are drawn from any nonsingular distribution~\cite{DSF94}.

\begin{figure}
\includegraphics[width=8cm]{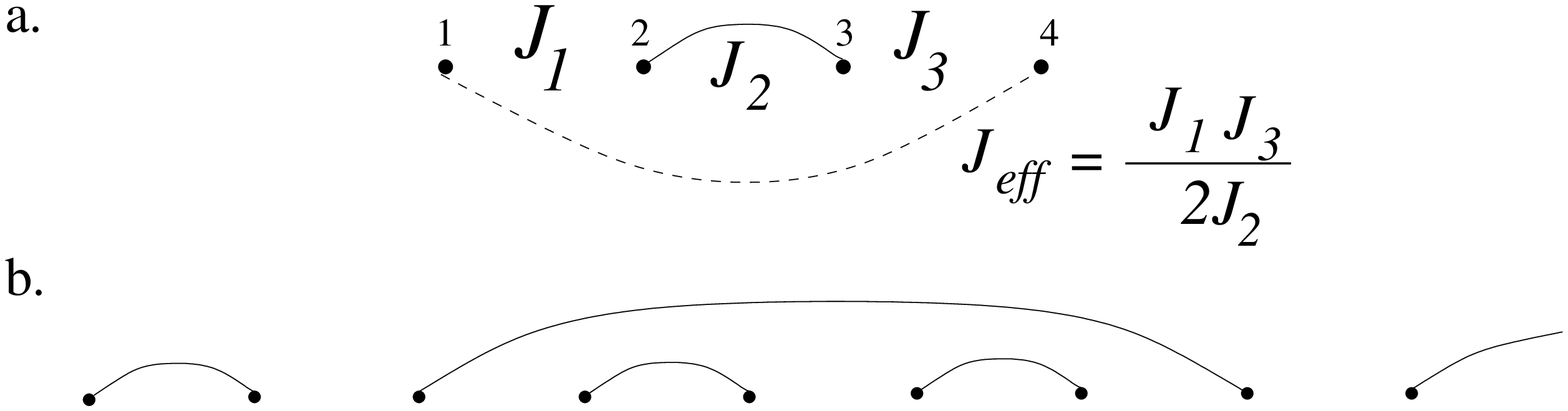}
\caption{a. RG decimation step: if $J_i$ is the strongest bond in the chain, sites $i$ and
  $i+1$ form a singlet (solid line); which diagonalizes $J_i {\bf
  S}_i \cdot {\bf S}_{i+1}$. Quantum fluctuations produce
  an effective interaction between sites $i-1$ and $i+2$
  (dashed line).
b. The random singlet ground state. Note that singlets may connect arbitrarily distant sites. 
\label{RSfig1}} 
\end{figure}

The real-space RG analysis consists of iteratively finding the
strongest bond, e.g. $J_i$, and diagonalizing it
independently of the rest of the chain. This leads to a singlet
between spins $i$ and $i+1$ in zeroth order (Fig \ref{RSfig1}a):
\be
|\psi^{(0)}\rangle=|\psi_{x<i}\rangle\frac{1}{\sqrt{2}}\l(\l|\uparrow_i\downarrow_{i+1}\rangle\r.-\l|\downarrow_{i}\uparrow_{i+1}\rangle\r.\r)|\psi_{x>i+1}\rangle.
\label{eq2.2}
\ee
Next, we treat the rest of the
Hamiltonian as a perturbation. If we begin with strong disorder (the
distribution of $\ln J_i$ is wide), we can assume that $J_i\gg J_{i-1},\,J_{i+1}$,
and use degenerate second order perturbation
theory. This leads to a Heisenberg interaction between the neighboring
spins at sites $i-1$ and $i+2$ with strength
\be
\tilde{J}_{i-1,\,i+2}=\frac{J_{i-1}J_{i+1}}{2J_i}<J_{i-1},\,J_{i},\,J_{i+1}.
\label{eq2.3}
\ee 
Thus we eliminate two sites, and reduce
the Hamiltonian's energy scale. Iterating these steps gives the ground state. Although this method is patently not correct when applied to a chain
with little disorder, it is still applicable and is
asymptotically correct at large distances.~\cite{DSF94}

The RG leads to an integro-differential flow equation for the
bond coupling distribution. This equation is best stated in terms of
the logarithmic coupling strength $\beta=\ln\frac{\Omega}{J}$ and RG flow parameter $\G=\ln \frac{\Omega_0}{\Omega}$, where
$\Omega_0$ is the Hamiltonian's initial
energy scale, and $\Omega$ is its reduced energy
scale. These variables capture the scaling properties of the problem;
e.g., neglecting a $\ln 2$, Eq. (\ref{eq2.3}) is simply
$\tilde{\beta}_{i-1,\,i+2}=\beta_{i-1}+\beta_{i+1}$. Note that strongest bonds have $\beta=0$. In terms of $\beta$ and $\G$ we have\cite{DSF94}
\beqn
\frac{dP_{\G}(\beta)}{d\G}&=&P_{\G}(0)\int d\beta_1 d \beta_2 \delta(\beta_1+\beta_2-\beta)P_{\G}(\beta_2)P_{\G}(\beta_1)
\cr&&+\frac{\partial P_{\G}(\beta)}{\partial
  \beta}.
\label{eq2.4}
\eeqn
The following solution is
an attractor to essentially all initial bond distributions, and it
describes the low-energy behavior of the spin chain:
\cite{DSF94}
\be
P(\beta)=\frac{1}{\G}e^{-\beta/\G}.
\label{eq2.5}
\ee
This is the random-singlet fixed
point distribution. 


The real space RG shows that the spin chain is in the random-singlet
phase. In this phase singlets form in a random fashion over all length
scales, and can connect spins arbitrarily far apart (Fig. \ref{RSfig1}b). Long-distance singlets form
at low energy scales. On average,
\be
\ra\sim\G^2,
\label{eq2.6}
\ee
where $\ra$ is the length scale of singlets forming at energy scale
$\Omega=\Omega_0 E^{-\G}$.
The long range of the low-energy singlets leads to slowly decaying
average correlations, which for the random Heisenberg model
decay algebraically, and not exponentially as one would expect from the
localized nature of the random-singlet state. 

\begin{figure}
\includegraphics[width=8cm]{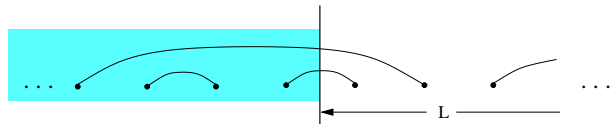}
\caption{The entanglement entropy of a segment is the
  number of singlets that connect the segment with the rest of the
  chain (shaded area). In this example there are two such singlets.
\label{RSfig2}} 
\end{figure}

Let us focus on the
entanglement entropy of the random Heisenberg model. The entanglement of a
spin-$1/2$ particle in a singlet with another such particle is $1$,
which is the entropy of the two states of a spin with its partner traced out. The entanglement of a segment of the random
Heisenberg chain is just {\it the number of singlets that
  connect sites inside to sites outside the segment} (Fig. \ref{RSfig2}). 

To obtain the entanglement, we calculate the number, $N$, of singlets that form over a single bond $B$ (in the example above, we
form singlets between sites $i$ and $i+1$, and later in the RG between
sites $i-1$ and $i+2$, etc.). If (as a first approximation) we neglect the history
dependence of the distribution of bond $B$, we can find $N$ by using
the distribution of bond strengths, Eq. (\ref{eq2.5}). When we
change the energy scale $\Omega\rightarrow \Omega-d\Omega$,
$\G\rightarrow \G+d\G$, all bonds with $0<\beta<d\G$ get
decimated.  The average number of decimations over bond $B$ grows by
\be
d\overline{N}=d\G P(\beta=0)=\frac{d\G}{\G}
\label{eq3.1}
\ee
which leads to $\overline{N}=\ln \G$.

This picture breaks down once singlets
exceed the size $L$ of the segment , when (by Eq. \ref{eq2.6})
$\G=\sqrt{L}$. So the number of singlets emanating from the segment of size $L$, i.e. its entanglement with the rest of the chain, is
\be
S_L\sim N_L\approx 2\cdot\ln \sqrt{L}+k=\ln L+k,
\label{eq3.3}
\ee
where $k$ is a nonuniversal constant, which also depends on the
initial realization of the disorder.


Neglecting the history of $B$ allowed us to see simply why the
entropy depends on $\ln L$, but the coefficient we obtained is not correct. To include the history of $B$, we note from
Eqs. (\ref{eq2.3}, \ref{eq2.5}) that the bond strength distribution of
$B$ right after being decimated at $\G_0$ is
\be
Q(\beta)=\int d\beta_1 d \beta_2
\delta(\beta_1+\beta_2-\beta)P_{\G_0}(\beta_2)P_{\G_0}(\beta_1)=\frac{\beta}{\G_0^2}e^{-\beta/\G}.
\label{eq3.4}
\ee
Now we ask at what $\G$ is $B$  decimated again. To answer this,
we need a flow equation for $Q_{\G}(\beta)$ similar to Eq. (\ref{eq2.4}).
In the following, we use the convention that  
$\int\limits_0^{\infty} d\beta Q_{\G}(\beta)=p_{\G}$ is the
probability that bond $B$ was not yet decimated at scale $\G$. With this
convention in mind, $Q_{\G}(\beta)$ obeys
\be
\ba{c}
\frac{dQ_{\G}(\beta)}{d\G}=\frac{\partial Q_{\G}(\beta)}{\partial
  \beta}-2Q_{\G}(\beta)P_{\G}(0)\vspace{2mm}\\
+2 P_{\G}(0)\int d\beta_1 d \beta_2 \delta(\beta_1+\beta_2-\beta)P_{\G}(\beta_2)Q_{\G}(\beta_1).
\label{eq3.5}
\ea
\ee
The first term is due to the change in $\beta$ when $\Omega$ changes,
the second and third terms account for $B$'s flow due to one of its two
neighbors forming a singlet. Note that
$\frac{dp_{\G}}{d\G}=-Q_{\G}(0)$. Eq. (\ref{eq3.5}) can be solved
using the ansatz:
\be
Q_{\G}(\beta)=\l(a_{\G}+b_{\G}\frac{\beta}{\G}\r)P_{\G}(\beta)
\label{eq3.6}
\ee
by substituting Eq. (\ref{eq3.6}) in Eq. (\ref{eq3.5}) we obtain
\be
\ba{cc}
\frac{da_{\G}}{dl}=\G\frac{da_{\G}}{d\G}=b_{\G}-2a_{\G}, &
\frac{db_{\G}}{dl}=\G\frac{db_{\G}}{d\G}=-b_{\G}+a_{\G}, 
\ea
\label{eq3.7}
\ee
with $l=\ln \G/\G_0$. Also $a_{\G_0}=0,\,b_{\G_0}=1$, from
Eq. (\ref{eq3.4}). 

Next we calculate the rate of singlet formation over 
$B$. First, note that the survival
probability $p_{\G}$ obeys
$p_{\G}=a_{\G}+b_{\G}$
and depends on $\G$ {\it only through} $l=\ln \G/\G_0$. Therefore the
duration $l$ between two consecutive singlets forming on bond $B$ is
{\it independent of} ~$\G_0$. This also proves that the number of
singlets over $B$ is proportional to $\ln \G$. To
find the proportionality constant we calculate the average duration
$\langle l\rangle$ between decimations; the number of bonds is then $N= \frac{\ln\G}{\langle l\rangle}$. 
We have
\be
\langle l\rangle=\int\,dp_{\G}\,l=\int\limits_0^{\infty}\,dl\,a_{\G} l.
\label{eq3.9}
\ee
From Eq. (\ref{eq3.7}) one finds $a_{\G}=\frac{1}{\sqrt{5}}\l(e^{-\frac{3-\sqrt{5}}{2}l}-e^{-\frac{3+\sqrt{5}}{2}l}\r).$
Inserting this in Eq. (\ref{eq3.9}) we find $\langle l\rangle=3$. Therefore:
\be
S_L=\frac{1}{3}\cdot 2 \ln\G+k=\frac{\ln 2}{3}\log_2 L+k.
\label{eq3.11}
\ee
Hence the 'effective central charge' of the random
Heisenberg chain is ${\tilde c}=1\cdot \ln 2$, which is the central
charge of the pure Heisenberg chain times an irrational number.


We discuss the interpretation and physical consequences of this effective central charge below, but first obtain its value for the quantum Ising case.  The pure quantum Ising
model has a well known ferromagnetic to paramagnetic phase transition. Its random analog is
\be
H = -\sum_i J_i \sz_i \sz_{i+1}-\sum_i h_i \sx_i, 
\label{QImodel}
\ee
where $h_i$ and $J_i$ are random, and $\sigma$ are spin-1/2 Pauli matrices.  This model also has a ferromagnetic to paramagnetic phase transition, described by a random critical point similar to the random singlet phase \cite{DSF95}.  Hence we expect entanglement in the quantum Ising case also to scale
logarithmically with the size $L$ of the test segment. 

\begin{figure}
\includegraphics[width=8cm]{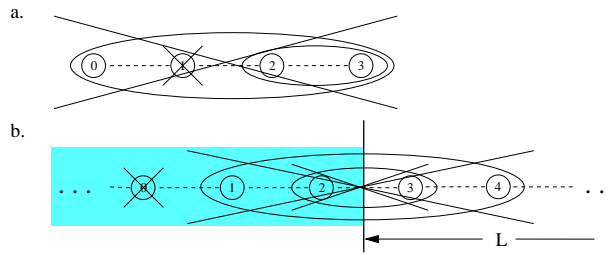}
\caption{(a) Typical
  ground state in the random quantum Ising model. It formed as follows; sites $2$ and $3$ form a cluster, site $1$ is
  frozen in the $x$ direction, site $0$ joins the cluster of $2$ and
  $3$, and finally the large cluster is frozen along $x$. 
(b) The entanglement of a segment $L$ is given by the
  number of decimated clusters that connect the segment with the rest of the
  chain (shaded area). In this example there are two such clusters:
  sites $2$ and $3$, and sites $1$ and $4$.
\label{RSfig3}} 
\end{figure}

As with the random Heisenberg case, we use real-space
RG to study the ground state of the random quantum Ising model (Fig. \ref{RSfig3}a). Again
we diagonalize the term with the largest energy scale in the Hamiltonian; if
it is $-J_i\sz_i \sz_{i+1}$, we set sites
$i$ and $i+1$ to point in the same direction, 
$|\psi\rangle_{i,i+1}=\alpha\l|\uparrow_i\uparrow_{i+1}\rangle\r.+\beta\l|\downarrow_i\downarrow_{i+1}\r.\rangle$,
thus creating a ferromagnetic {\it cluster}. Quantum fluctuations
yield an {\it effective transverse field} on the
cluster,
\be
h_{i,i+1}=\frac{h_ih_{i+1}}{J_i}\,\,\l(\ll h_i,\,h_{i+1},\,J_i\r).
\ee
If the term with largest energy happens to be $-h_i\sx_i$, we set the $i$'th spin point in the $x$
direction,
$|\psi\rangle_i=\l|\rightarrow_i\rangle\r.=\frac{1}{\sqrt{2}}\l(\l|\uparrow_i\rangle\r.+\l|\downarrow_i\rangle\r.\r)$,
by which we decimate the $i$'th spin. Quantum fluctuations
produce an
effective Ising coupling between sites $i-1$ and $i+1$ with strength:
\be
J_{i-1,i+1}=\frac{J_{i-1}J_i}{h_i}\,\,\l(\ll J_{i-1},\,J_{i},\,h_i\r).
\ee

The RG flow equations for the distributions
of $h_i$ and $J_i$ as a function of $\Omega=\max\{J_i,\,h_i\}$ support an attractor in which the logarithmic coupling distributions, $R(\zeta)$
and $P(\beta)$, with  
$\zeta=\ln\frac{\Omega}{J}$, $\beta=\ln\frac{\Omega}{h}$, are given by
the random singlet expression in
Eq. (\ref{eq2.5}).  As for the random singlet scaling, at criticality
the absolute length $\ra$ of the domains decimated by the
transverse field scales as $\ra\sim\Gamma^2$.


At low energy, larger and larger ferromagnetic clusters are formed and then
decimated. For a segment of length $L$, ferromagnetic clusters
which are completely within or completely outside the segment and decimated by the transverse field do not
affect the entanglement. The only contributions come from ferromagnetic
domains that cross the boundary of the segment
(Fig.~\ref{RSfig3}b), and each such
cluster contributes $1$ to the entanglement.


Next we calculate how many ferromagnetic clusters are
formed and decimated over an edge of the segment.
At a given energy scale, the edge of the segment can either separate two unpaired sites, or be contained in a cluster which is
partially in the segment (when such a cluster is decimated the
edge returns to the first case). At
the critical point these possibilities must occur with the same
probability by a self-duality of the quantum Ising model
(\cite{DSF95,Refael2003}). Hence the probability that the edge is
in an active cluster at scale $\Gamma$ is $p=1/2$.
By
the analysis as for the Heisenberg chain, we obtain
$N(\Gamma)=p\cdot\frac{1}{3}\ln\Gamma$, and
\be
S_L=\frac{1}{6}\ln L+k=\frac{\ln 2}{6}\log_2 L+k,
\label{QIE}
\ee
with $k$ a non-universal constant.
The effective
central charge of the random quantum Ising model is ${\tilde
  c}=1/2\cdot \ln 2$ - $\ln 2$ times the central charge in the pure system.


Eq. (\ref{QIE}) shows that the critical quantum Ising chain has half the entanglement 
of the random singlet phase. But both of these are in the same infinite-randomness fixed point scaling category, so a difference in the entanglement entropy may seem surprising. However, these two systems also differ in their temporal correlations:
\be
\overline{<\sz_0(0)\sz_0(\tau)>}\sim\frac{1}{\ln^z{\Omega_0 \tau}},
\ee    
with $z=2$ for the XXZ models and $z=1$ for the quantum Ising model.
These two similar strong-randomness fixed points belong to different universality classes, and the effective central charge is sensitive to this difference.



For pure chains, the prefactors of correlation functions are
non-universal but the central charge is universal. In the random case,
non-universal correlation prefactors are generated by inaccuracies
of the order of the lattice spacing in the location of the low energy
effective spins; these occur when the RG scale $\Omega$ is still large\cite{DSF98}.  Such errors do not affect the universal
coefficient of the logarithmic divergence of the entropy, but modify the additive non-universal 'surface term' $k$ in Eqs. (\ref{eq3.11}, \ref{eq3.3}).
Note that entanglement is self-averaging as chain length $N \rightarrow \infty$.

  
The central charges we find for the random Heisenberg, XX, and quantum Ising
chains are those of the pure models times $\ln 2$. Although 
irrational, these central charges are universal quantities which
describe the universality class of the {\it random} chains. An example of the importance of $c$ in the pure case is the well known
``$c$-theorem''~\cite{zamolodchikov} that if an RG flow connects one
critical point $A$ to another critical point $B$, then $c_A \geq c_B$.
The values of ${\tilde c}$ obtained here for random systems suggest,
given the relevance of disorder in these systems, that there is a generalized ${\tilde c}$-theorem
based on entanglement even for nonconformal quantum critical points in
1D.  This may imply constraints on the values of ${\tilde c}$ for
spin chains obtained by, e.g., disordering higher-spin CFTs. 

Since only certain rational values of $c$ are allowed for well-behaved
CFTs with $c<1$, the irrational ${\tilde c}$
for random critical points is a fundamental difference between
pure and random cases.  This is reminiscent of the irrational
residual entropy that appears in quantum impurity problems and
satisfies a ``$g$-theorem.'' \cite{AffleckLudwig}

The ratio between the random and pure values of ${\tilde c}$ is
unexpectedly the same for all the different chains we studied.
Perhaps $\tilde c$ of any {\it random} fixed point derived from a
pure conformally invariant point is the product of the
central charge of the pure theory and a universal number determined by
the flow from the pure to the random fixed point.
Numerics on the random $XX$ model via its free-fermion representation
could determine whether ${\tilde c}$ appears as a universal
amplitude as in the clean case for quantities beyond entanglement.  It
is clear already that the universal logarithmic scaling of
entanglement provides a powerful way to characterize both pure and
random quantum critical points in one dimension.

We gratefully acknowledge useful conversations with L. Balents, A. Kitaev,
A. W. W. Ludwig, J. Preskill, and G. Vidal, and support from NSF PHY99-07949, DMR-0238760, and the Hellman Foundation.

\bibliography{bigbib}

\end{document}